\documentclass[3p,times,twocolumn]{elsarticle}

\usepackage{ecrc}

\volume{00}

\firstpage{1}

\journalname{Nuclear Physics B Proceedings Supplement}

\runauth{}

\jid{nuphbp}

\jnltitlelogo{Nuclear Physics B Proceedings Supplement}

\usepackage{amssymb}
\usepackage[figuresright]{rotating}

\def\bpm{\begin{pmatrix}}
\def\epm{\end{pmatrix}} 
\def\d{{\rm d}}

\def\st{\tilde\tau}
\def\thst{\theta_{\st}}
\def\mgl{m_{\tilde g}}
\def\msq{m_{\tilde q}}
\newcommand{\neu}[1]{\tilde \chi^0_{#1}}
\def\tb{\tan\beta}

\def\lr{\left( }
\def\rr{\right) }
\def\le{\left[ }
\def\re{\right] }

\def\beq{\begin{equation}}
\def\eeq{\end{equation}}
\def\bea{\begin{eqnarray}}
\def\eea{\end{eqnarray}}

\begin{document}

\begin{frontmatter}



\dochead{}

\title{Precision predictions for direct gaugino and slepton production at the LHC}


\author{B. Fuks}

\address{CERN, PH-TH, CH-1211 Geneva 23, Switzerland; \\
  Institut Pluridisciplinaire Hubert Curien/D\'epartement Recherches
  Subatomiques, Universit\'e de Strasbourg/CNRS-IN2P3, 23 Rue du Loess, F-67037
  Strasbourg, France}

\author{M. Klasen, D.R. Lamprea, M. Rothering}

\address{Institut f\"ur Theoretische Physik, Westf\"alische
 Wilhelms-Universit\"at M\"unster, Wilhelm-Klemm-Stra\ss{}e 9,
 D-48149 M\"unster, Germany}

\begin{abstract}
The search for electroweak superpartners has recently moved to the centre of interest at the
LHC. We provide the currently most precise theoretical predictions for these particles,
use them to assess the precision of parton shower simulations, and reanalyse public
experimental results assuming more general decompositions of gauginos and sleptons.
\end{abstract}


\end{frontmatter}

% \linenumbers


\vspace*{-15cm}
CERN-PH-TH-2014-144, MS-TP-14-26
\vspace*{13.4cm}

\section{Introduction}
\label{}

For many theoretical and phenomenological reasons, supersymmetry (SUSY) remains one of the best
motivated extensions of the Standard Model (SM) of particle physics. The strongly
interacting superpartners in the Minimal SUSY SM (MSSM), the first- and second generation
squarks and gluinos, are largely restricted after the first LHC run at 7 and 8 TeV
centre-of-mass energy to be heavier than 1 TeV. However, this is not the case for stops,
which play a
central role in explaining the relatively large mass of the SM-like Higgs boson, and the
electroweakly interacting sleptons and gauginos, which provide natural candidates for the dark
matter in the universe. The search for these particles has therefore recently moved to the
centre of interest at the LHC.

LHC analyses on SUSY particle searches rely heavily on precision calculations of SM backgrounds
and SUSY signals. At next-to-leading order (NLO) of QCD, SUSY production cross sections have
been calculated more than a decade ago \cite{Beenakker:1996ch,Beenakker:1997ut,Berger:1998kh,%
Beenakker:1999xh,Berger:1999mc,Berger:2000iu,Spira:2002rd}. More recently, resummation methods
have been applied at next-to-leading logarithmic (NLL) accuracy \cite{Kramer:2012bx}.
Here, we present our NLO+NLL calculations for direct gaugino \cite{Debove:2009ia,Debove:2010kf,%
Debove:2011xj,Fuks:2012qx} and slepton \cite{Bozzi:2006fw,Bozzi:2007qr,Bozzi:2007tea,%
Fuks:2013lya} production near threshold and close to vanishing transverse momentum ($p_T$), use
them to assess the precision of parton shower simulations, and reanalyse public experimental
results assuming more general decompositions of gauginos and sleptons.

\section{Resummation}
\label{}

The hadronic cross section for the production of SUSY particles at the LHC
\bea
 \sigma_{pp}&=&f_{a/p}(x_a,\mu_f)\otimes f_{b/p}(x_b,\mu_f)\otimes \nonumber\\ &&
 \sum_{n=0}^\infty \alpha_s^n(\mu_r) \sigma_{ab}^{(n)}(\mu_r,\mu_f)
\eea
is obtained by a convolution of the parton densities (PDFs) $f(x,\mu_f)$, that depend
on the partonic momentum fraction $x$ and the factorisation scale $\mu_f$, with the partonic
cross section $\sigma_{ab}$, that can be expanded in powers of the strong coupling constant
$\alpha_s(\mu_r)$ running with the renormalisation scale $\mu_r$.

Near production threshold, where the ratio of the squared invariant mass $M^2$ of the produced
particle pair over the partonic centre-of-mass energy $s$, $z={M^2/s}$, approaches unity,
the cross section exhibits logarithimc enhancements,
\beq
 \sigma^{(n)}_{q\bar{q}}(z)=\sum_{m=0}^{2n-1}c^{(m)}\le{\ln^m(1-z)\over(1-z)}\re_+.
\eeq
After applying a Mellin transform, e.g. 
\beq
 \le{\ln^m(1-z)\over(1-z)}\re_+\to \ln^{m+1}(N),
\eeq
these logarithmically enhanced terms, coming from soft gluon radiation,
can be resummed to all orders,
\beq
 \sigma^{(n)}_{q\bar{q}}(N)=H_{q\bar{q}}\cdot\exp\lr \tilde{c}^{(1)}\ln(N)+\tilde{c}^{(2)}+...\rr,
\eeq
where $H$ represents the hard, non-singular part and $\tilde{c}^{(i)}$ are universal coefficients.
Since also the dominant collinear $1/N_C$ terms ($N_C$ being the number of colours in QCD)
are universal, they can also be exponentiated in a so-called ``collinear improved'' resummation
calculation \cite{Kramer:1996iq}.

A second critical region is encountered when the transverse momentum of the produced particle
pair tends to zero, $p_T\to0$. There, the cross section behaves as
\beq
 \sigma^{(n)}_{q\bar{q}}(p_T)=\sum_{m=0}^{2n-1}c^{(m)}\le{1\over p_T^2}\ln^m \lr{M^2\over p_T^2}\rr\re_+.
\eeq
After applying a Fourier transform,
\beq
 \le{1\over p_T^2}\ln^m\lr{ M^2\over p_T^2}\rr\re_+ \to \ln^{m+1}\lr{M^2b^2\over b_0^2}+1\rr
\eeq
with $b_0=2e^{-\gamma_E}$, the logarithms can again be resummed  to all orders,
\beq
 \sigma^{(n)}_{q\bar{q}}(N)=H_{q\bar{q}}\cdot\exp\lr \tilde{c}^{(1)}\ln(b)+\tilde{c}^{(2)}+...\rr.
\eeq
Since the threshold and transverse-momentum logarithms are of the same kinematic origin,
i.e.\ soft gluon radiation, they can also be resummed jointly in ($N,b$) space.

To achieve the best possible accuracy over the full kinematic ranges, the fixed-order and
resummed results are added. However, since the logarithmically enhanced terms are present in both
parts, this overlap must be subtracted to avoid double counting,
\beq
 \sigma_{ab}=\sigma_{ab}^{\rm f.o.}+\sigma_{ab}^{\rm res.}-\sigma_{ab}^{\rm exp.}.
\eeq
Distributions in the measured quantities $M$ and $p_T$ are then obtained by applying
an inverse Mellin transform
\beq
 M^2{\d\sigma_{AB}(\tau)\over\d M^2}={1\over2\pi i}\int_{{\cal C}_N}\d N
 \tau^{-N}M^2{\d\sigma_{AB}(N)\over\d M^2}
\eeq
with the so-called minimal prescription, where an integration contour
$ {\cal C}_N$ is defined by $N=C+ze^{\pm i\phi}$ and $ z\in[0;\infty[$,
and an inverse Fourier transform
\beq
{\d\sigma\over\d p_T^2}={M^2\over s}\int_0^\infty \d b \
{b\over2} \ J_0(bp_T) \ \d\sigma(b)
\eeq
 with a deformed contour  $ b=(\cos\phi+i\sin\phi)t$ and
$ t\in[0;\infty[$ for a proper treatment of all encountered poles in the complex plane.

\section{Gauginos}
\label{}

Using the resummation formalisms described briefly above, we demonstrate the impact of our
precision predictions for gaugino pair production at the LHC with a centre-of-mass energy of
$\sqrt{s}=8$ TeV at the constrained MSSM benchmark point defined in Tab.\ \ref{tab:1}.
%
%%%%%%%%%%%%%% Begin Table 1 %%%%%%%%%%%%%%%%%%%%%%%%%%%%%%%%%%%%%%%%%%%
\begin{table}[t]
\caption{\label{tab:1}
 Our constrained MSSM benchmark point with $\tb = 10$ and $A_0 = 0$ GeV.
 All masses are given in units of GeV, and the gluino and average squark masses are
 rounded to 5 GeV accuracy.}
\begin{center}
\begin{tabular}{|c||c|c||c|}
\hline
 $(m_{1/2}, m_0)$ & $\mgl$ & $\langle \msq \rangle$
 & BR$(\neu{2} \to \neu{1} h)$ \\
\hline
 (600, 400) & 1370 & 1275 & 92\% \\
\hline
\end{tabular}
\end{center}
\end{table}
%%%%%%%%%%%%%% End of Table 1 %%%%%%%%%%%%%%%%%%%%%%%%%%%%%%%%%%%%%%%%%%
%
It features sufficiently high squark and gluino masses, that are not yet excluded,
and an interestingly large branching fraction of the second lightest neutralino into
the lightest neutralino
and the SM-like Higgs boson. The neutralino/chargino masses are 250 GeV
for $\tilde\chi_1^0$, 472 GeV for $\tilde\chi_2^0/\tilde\chi_1^{\pm}$ and 766 for $\tilde\chi_{3,4}^0/\tilde\chi_2^{\pm}$,
and the corresponding total cross sections are shown in Tab.\ \ref{tab:2}.
%
%%%%%%%%%%%%%% Begin Table 2 %%%%%%%%%%%%%%%%%%%%%%%%%%%%%%%%%%%%%%%%%%%
\begin{table}[!t]
\caption{\label{tab:2}
 Total cross sections (in fb) for the production of various gaugino pairs and their
 associated scale and PDF uncertainties for the LHC running at a center-of-mass energy
 of $\sqrt{s}=8$ TeV at our benchmark point.  The PDF uncertainties are not shown for
 the LO results.}
\renewcommand{\arraystretch}{1.2}
\begin{center}
\begin{tabular}{| c || l | l | l |}
\hline
 $\tilde\chi_i\tilde\chi_j$ &  LO & NLO  & NLO+NLL \\
\hline
$\tilde\chi^0_1 \tilde\chi^0_1$ & $0.13^{+8.6 \%}_{-7.5 \%}$  &  $0.16^{+3.5 \%}_{-3.4 \%}{}^{+3.3 \%}_{-2.3 \%}$  & $0.16^{+0.2 \%}_{-0.3 \%}{}^{+3.5 \%}_{-2.4 \%}$\\
$\tilde\chi^0_2 \tilde\chi^-_1$ & $1.63^{+10.0 \%}_{-8.6 \%}$  &  $1.88^{+1.8 \%}_{-2.4 \%}{}^{+4.1 \%}_{-3.1 \%}$  & $1.86^{+0.6 \%}_{-1.2 \%}{}^{+4.1 \%}_{-3.1 \%}$\\
$\tilde\chi^+_1 \tilde\chi^0_2$ & $4.73^{+9.8 \%}_{-8.4 \%}$  &  $5.28^{+1.8 \%}_{-2.4 \%}{}^{+3.9 \%}_{-2.5 \%}$  & $5.22^{+0.3 \%}_{-0.6 \%}{}^{+4.0 \%}_{-2.5 \%}$\\
$\tilde\chi^+_1 \tilde\chi^-_1$ & $3.13^{+9.8 \%}_{-8.4 \%}$  &  $3.57^{+1.9 \%}_{-2.5 \%}{}^{+3.5 \%}_{-2.2 \%}$  & $3.52^{+0.4 \%}_{-0.7 \%}{}^{+3.7 \%}_{-2.3 \%}$\\
$\tilde\chi^+_2 \tilde\chi^0_3$ & $0.16^{+14.2 \%}_{-11.6 \%}$  &  $0.17^{+3.5 \%}_{-4.2 \%}{}^{+6.1 \%}_{-3.8 \%}$  & $0.17^{+1.0 \%}_{-1.8 \%}{}^{+6.1 \%}_{-3.8 \%}$\\
$\tilde\chi^+_2 \tilde\chi^0_4$ & $0.15^{+14.3 \%}_{-11.7 \%}$  &  $0.16^{+3.4 \%}_{-4.2 \%}{}^{+6.1 \%}_{-3.9 \%}$  & $0.16^{+1.1 \%}_{-1.8 \%}{}^{+6.1 \%}_{-4.0 \%}$\\
$\tilde\chi^+_2 \tilde\chi^-_2$ & $0.11^{+13.6 \%}_{-11.2 \%}$  &  $0.12^{+3.1 \%}_{-3.9 \%}{}^{+6.0 \%}_{-3.5 \%}$  & $0.12^{+1.0 \%}_{-1.8 \%}{}^{+6.0 \%}_{-3.6 \%}$\\
\hline
\end{tabular}
\end{center}
\end{table}
%%%%%%%%%%%%%% End of Table 2 %%%%%%%%%%%%%%%%%%%%%%%%%%%%%%%%%%%%%%%%%%
%
As one can see, they are often increased, in particular from LO to NLO and, as
one approaches the production threshold, also from NLO to NLL, and the scale
uncertainty is always considerably stabilised.

It is interesting to compare the NLL threshold resummed results with a Monte Carlo
prediction at LO using the multi-parton generator MadGraph \cite{Alwall:2011uj}
and the PYTHIA \cite{Sjostrand:2006za} parton shower.
%
%%%%%%%%%%%%%% Begin Figure 1 %%%%%%%%%%%%%%%%%%%%%%%%%%%%%%%%%%%%%%%%%%
\begin{figure}
 \centering
 \includegraphics[width=\columnwidth]{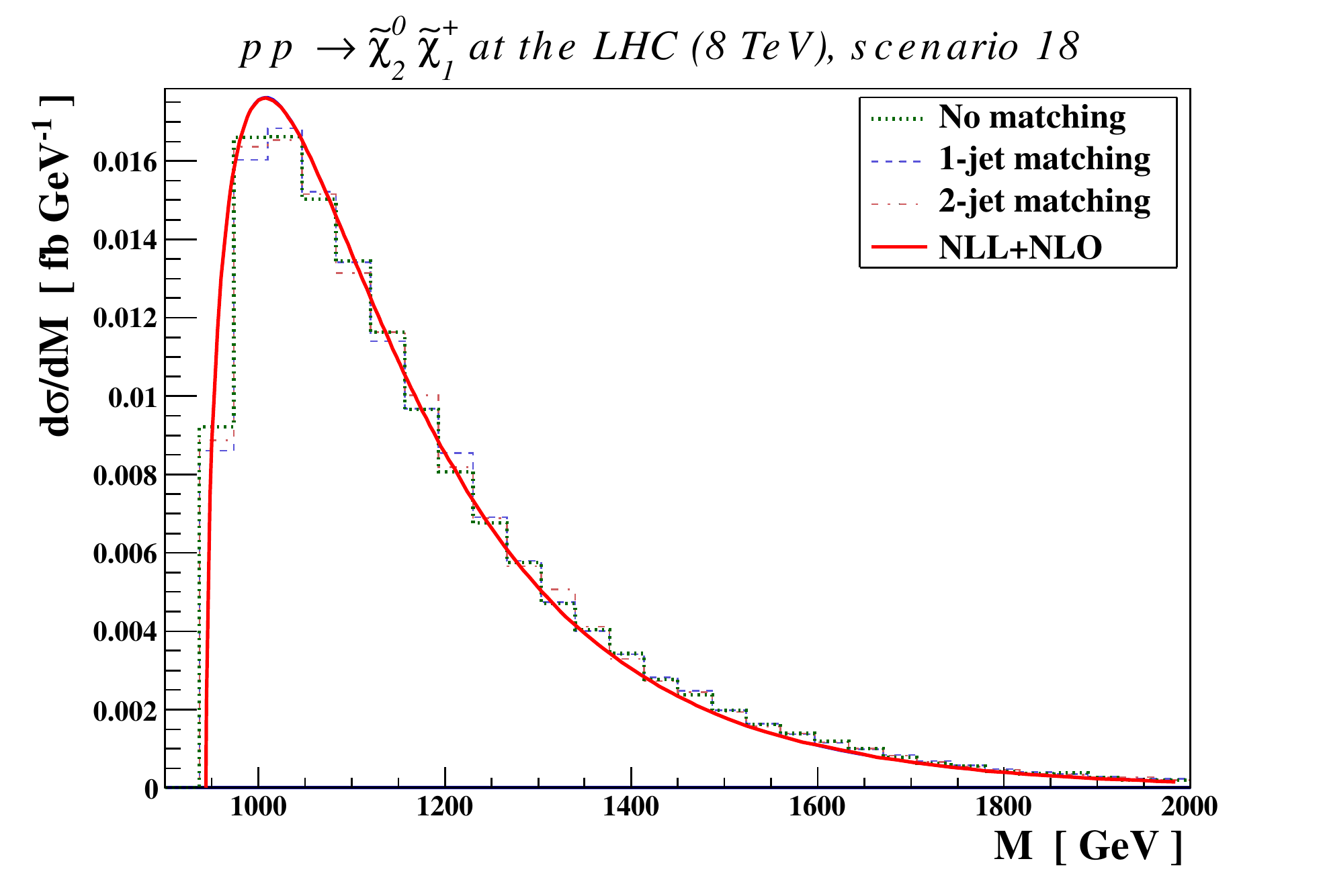}
 \caption{\label{fig:1}Distributions in the invariant mass $M$ of a
$\tilde \chi^0_2 \tilde\chi^+_1$ pair with mass 472 GeV each at the LHC
with $\sqrt{s}=8$ TeV. We compare 
the NLO matched to the NLL (red, thick full) distribution to the results
obtained after matching matrix elements containing no (green, dotted), one (blue,
dashed), and up to two (red, dot-dashed) additional jets to parton showering.}
\end{figure}
%%%%%%%%%%%%%% End of Figure 1 %%%%%%%%%%%%%%%%%%%%%%%%%%%%%%%%%%%%%%%%%
%
As one can see in Fig.\ \ref{fig:1}, the NLL+NLO invariant mass distribution (red, thick full)
agrees in general very well with the Monte Carlo results obtained after matching matrix elements
containing no (green, dotted), one (blue, dashed), and up to two (red, dot-dashed) additional
jets to parton showering. As the NLO+NLL calculation does not contain more than one hard
additional jet, it does, however, not allow to validate precisely the two-jet matching
\cite{Mangano:2002ea}.

\begin{figure}
 \centering
 \includegraphics[width=\columnwidth]{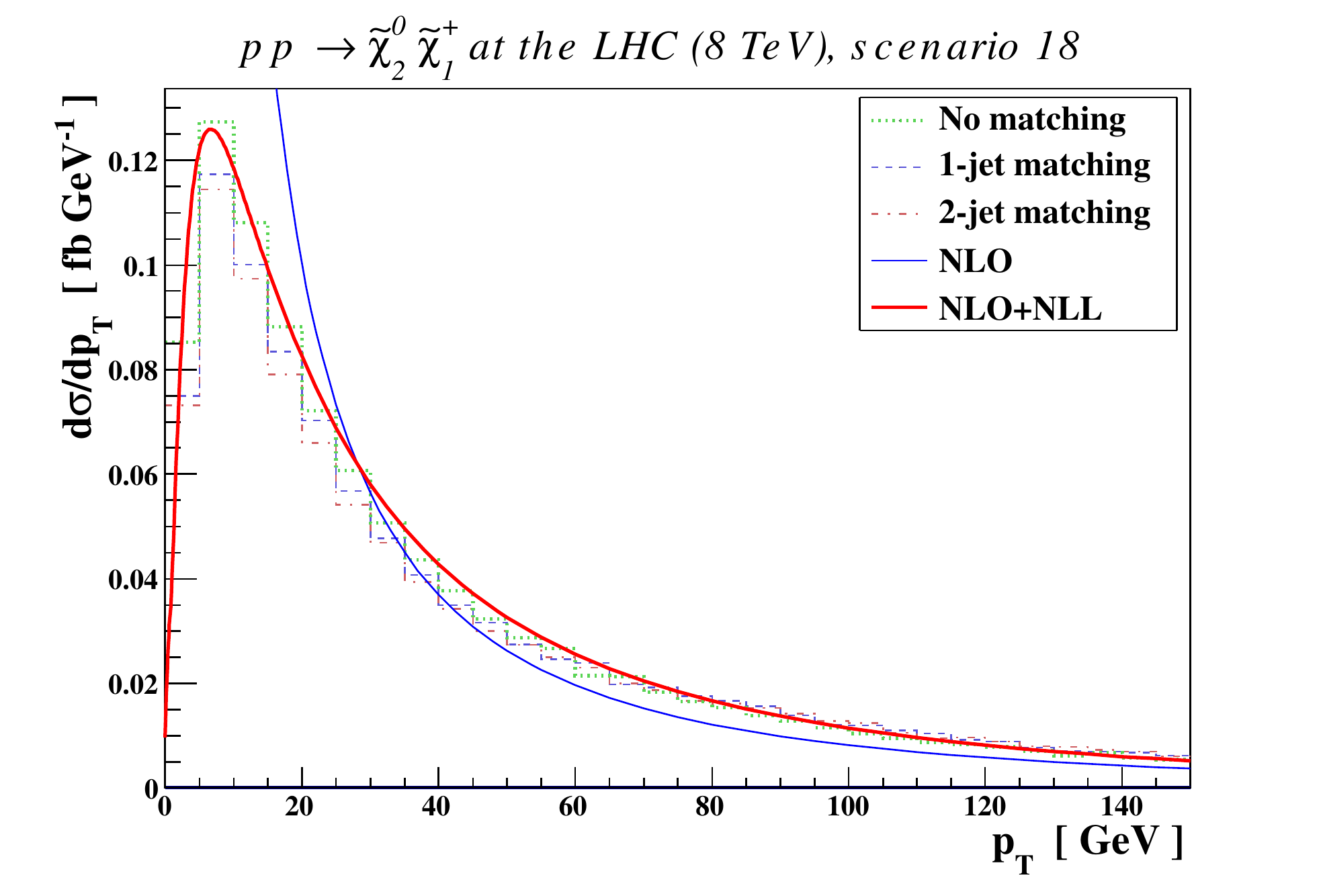}
 \caption{\label{fig:2}Distributions in the transverse momentum $p_T$ of a
$\tilde\chi^0_2\tilde\chi^+_1$ pair with mass 472 GeV each at the LHC
with $\sqrt{s}=8$ TeV. We compare fixed order at ${\cal O}(\alpha_s)$ (blue, full)
and NLL (red, thick full) distributions to the results
obtained after matching matrix-elements containing no (green, dotted), one (blue,
dashed), and up to two (red, dot-dashed) additional jets to parton showering.}
\end{figure}

A comparison of NLO and NLO+NLL $p_T$ spectra versus the corresponding MadGraph and PYTHIA
predictions is shown in Fig.\ \ref{fig:2}. While the NLO prediction diverges at low $p_T$,
%
%%%%%%%%%%%%%% Begin Figure 2 %%%%%%%%%%%%%%%%%%%%%%%%%%%%%%%%%%%%%%%%%%
%%%%%%%%%%%%%% End of Figure 2 %%%%%%%%%%%%%%%%%%%%%%%%%%%%%%%%%%%%%%%%%
%
the NLO+NLL result shows the correct physical turnover and agrees very well
with the Monte Carlo predictions. Again, the two-jet matching can not be
precisely validated due to the lack of two hard jets in the NLO+NLL calculation.

\section{Sleptons}
\label{}

The production of slepton ($\tilde{l}$) pairs has so far been analysed by the LHC experiments
ATLAS and CMS using simplified models. In particular, they assume a flavour-conserving decay
into a SM lepton $l$ and the lightest SUSY particle (LSP, $\tilde{\chi}^0_1$), while all other
SUSY particles, in particular the squarks and gluinos, are assumed to be heavy and to decouple.
The experimental signature is then a pair of same-flavour leptons and missing transverse
energy ($\not{\!E}_T$).

In our (re-)analysis \cite{Conte:2012fm},
we take into account different slepton flavors (also $\tilde{\tau}$),
both left- and right-handed sleptons (incl.\ mixing for staus) \cite{Bozzi:2004qq}, and a
different gaugino or higgsino nature of lightest neutralino \cite{Debove:2008nr}. The stau
mass eigenstates are in particular obtained through
\beq
\lr\begin{array}{c}\st_1\\ \st_2 \end{array}\rr = 
\lr\begin{array}{rr}\cos\thst & \sin\thst \\-\sin\thst & \cos\thst\end{array}\rr 
\lr\begin{array}{c}\st_L\\ \st_R \end{array}\rr.
\eeq
Their couplings to $Z$ bosons and neutralinos are given by
\bea
 \!\!\!\!\!\! {\cal C}^{(\tau)}_Z   \!&\!=\!&\! \Big[-\frac12 + s_W^2\Big] \cos^2\thst + \Big[  s_W^2  \Big] \sin^2\thst \\
 \!\!\!\!\!\! {\cal C}^{(\tau,L)}_N \!&\!=\!&\!\!\! \sqrt{2} e \Big[s_W N_1^\ast + c_W N_2^\ast \Big] \cos\thst
    - \Big[2 c_W s_W N_3^\ast y_\tau\Big] \sin\thst \nonumber\\
 \!\!\!\!\!\! {\cal C}^{(\tau,R)}_N \!&\!=\!&\! \Big[-2 \sqrt{2} e s_W N_1\Big] \sin\thst 
    - \Big[2 c_W s_W N_3 y_\tau\Big] \cos\thst\nonumber
\eea
where $y_\tau$ denotes the tau lepton Yukawa coupling, which in the case of third-generation
(s)leptons cannot be neglected. The four neutralino mixing parameters are constrained by
a unitarity relation,
\beq
  |N_1|^2 + |N_2|^2 + |N_3|^2 + |N_4|^2 =1 \ .
\eeq

\begin{figure}
 \centering
  \includegraphics[width=\columnwidth]{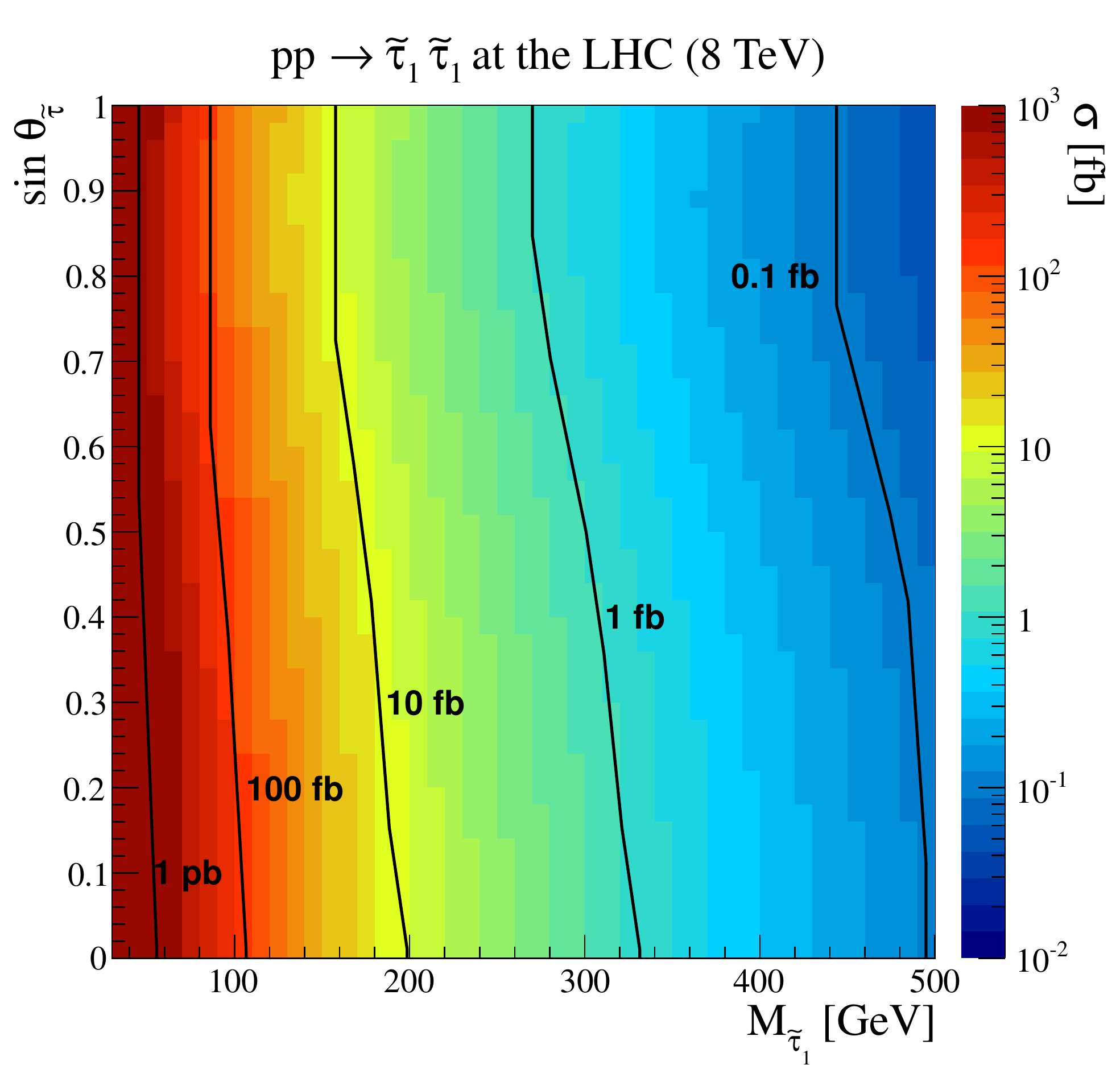}
\caption{\label{fig:3}Total cross sections for stau
  pair production at the LHC, running at
  centre-of-mass energies of 8~TeV. We present
  predictions as functions of the stau mass and the stau mixing
  angle after matching the NLO results with threshold resummation at the NLL
  accuracy.}
\end{figure}

In Fig.\ \ref{fig:3} we show total production cross sections at NLO+NLL as
%
%%%%%%%%%%%%%% Begin Figure 3 %%%%%%%%%%%%%%%%%%%%%%%%%%%%%%%%%%%%%%%%%%
%%%%%%%%%%%%%% End of Figure 3 %%%%%%%%%%%%%%%%%%%%%%%%%%%%%%%%%%%%%%%%%
%
a function of both stau mass and mixing angle. As one can see, the cross section
drops with the mass of the produced staus, but also as they become more right-handed,
corresponding to larger values of $\theta_{\tilde\tau}$.

\begin{figure}
 \centering
  \includegraphics[width=\columnwidth]{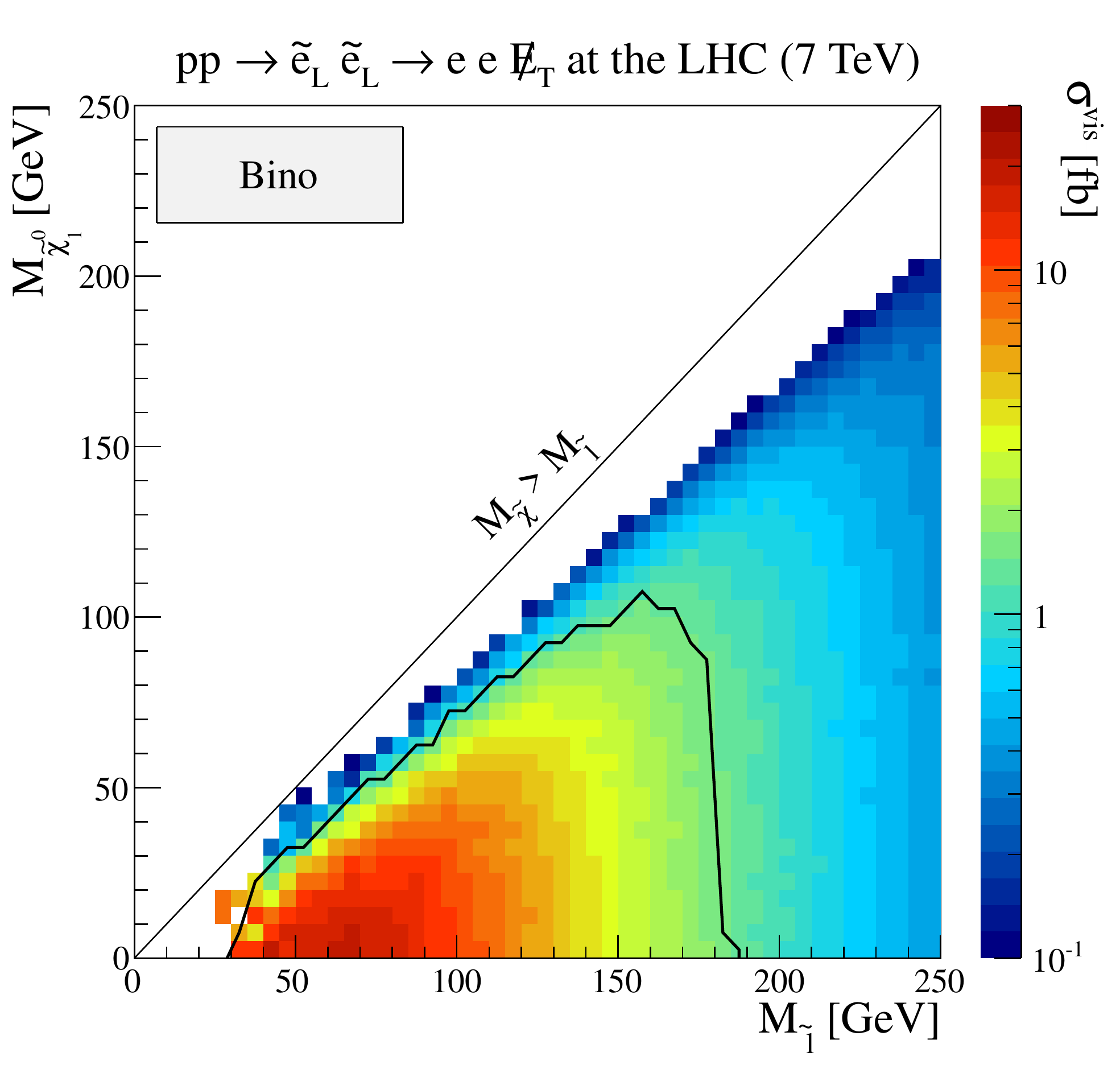}
  \includegraphics[width=\columnwidth]{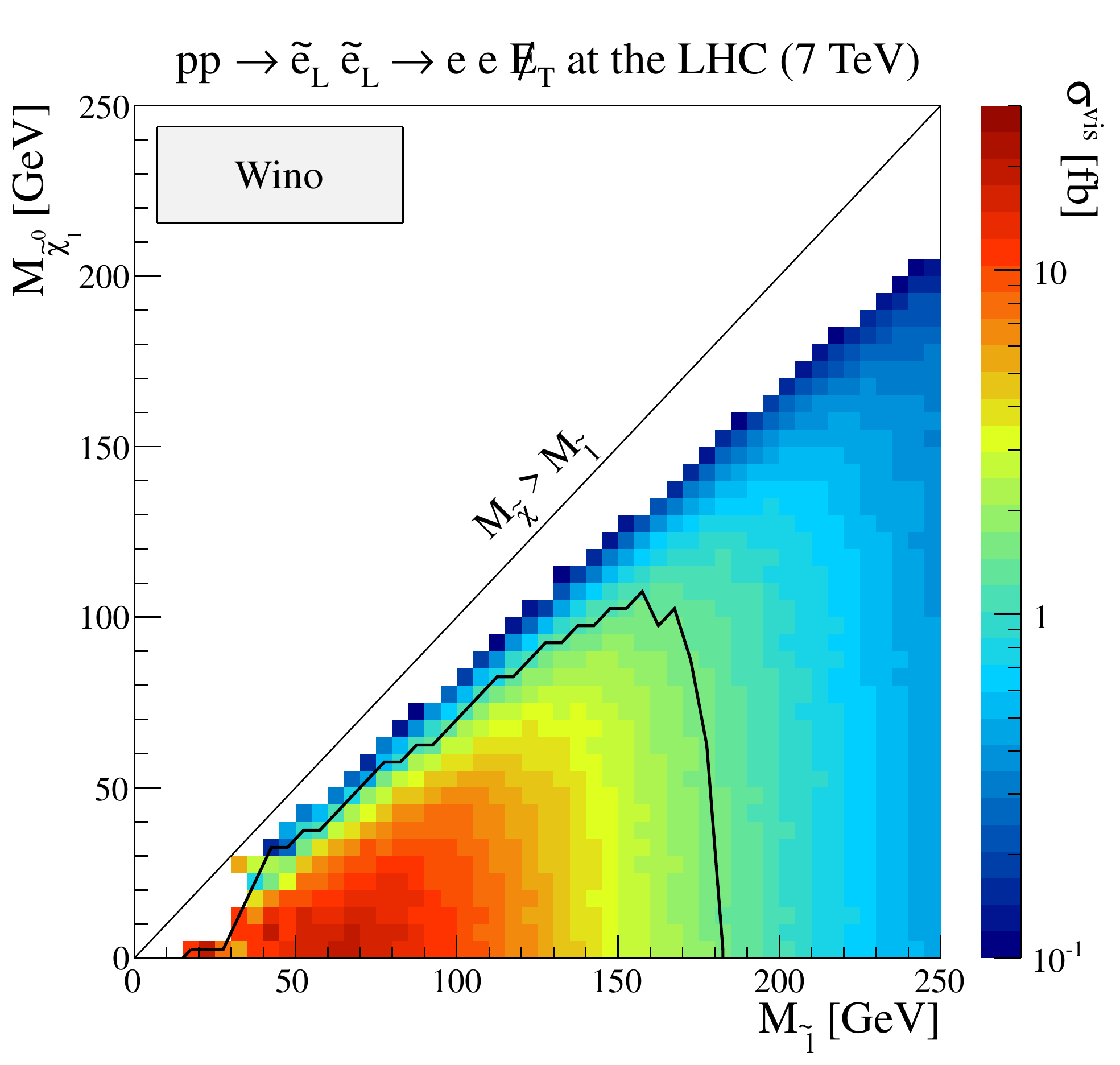}
\caption{\label{fig:4}95\% confidence exclusion limit for left-handed
selectron pair production,
given in the $(M_{\tilde{l}},M_{\tilde\chi^0_1})$ mass plane of a simplified
model for different choices of the neutralino nature
taken as bino (top) and wino (bottom). We present the visible cross
section after applying the ATLAS selection strategy.
The limits are extracted for 4.7~fb$^{-1}$ of LHC collisions at a centre-of-mass
energy of 7~TeV.}
\end{figure}

We have recast a recent ATLAS slepton analysis \cite{Aad:2012pxa} to take into account
the different gaugino/higgsino nature of the neutralino that results, e.g., from
left-handed selectron decays. As one can see in Fig.\ \ref{fig:4}, the exclusion curves
%
%%%%%%%%%%%%%% Begin Figure 4 %%%%%%%%%%%%%%%%%%%%%%%%%%%%%%%%%%%%%%%%%%
%%%%%%%%%%%%%% End of Figure 4 %%%%%%%%%%%%%%%%%%%%%%%%%%%%%%%%%%%%%%%%%
%
for binos (top) and winos (bottom) are very similar, i.e.\ there is not much
sensitivity to the nature of the lightest neutralino in these slepton decays.

\begin{figure}
 \centering
  \includegraphics[width=\columnwidth]{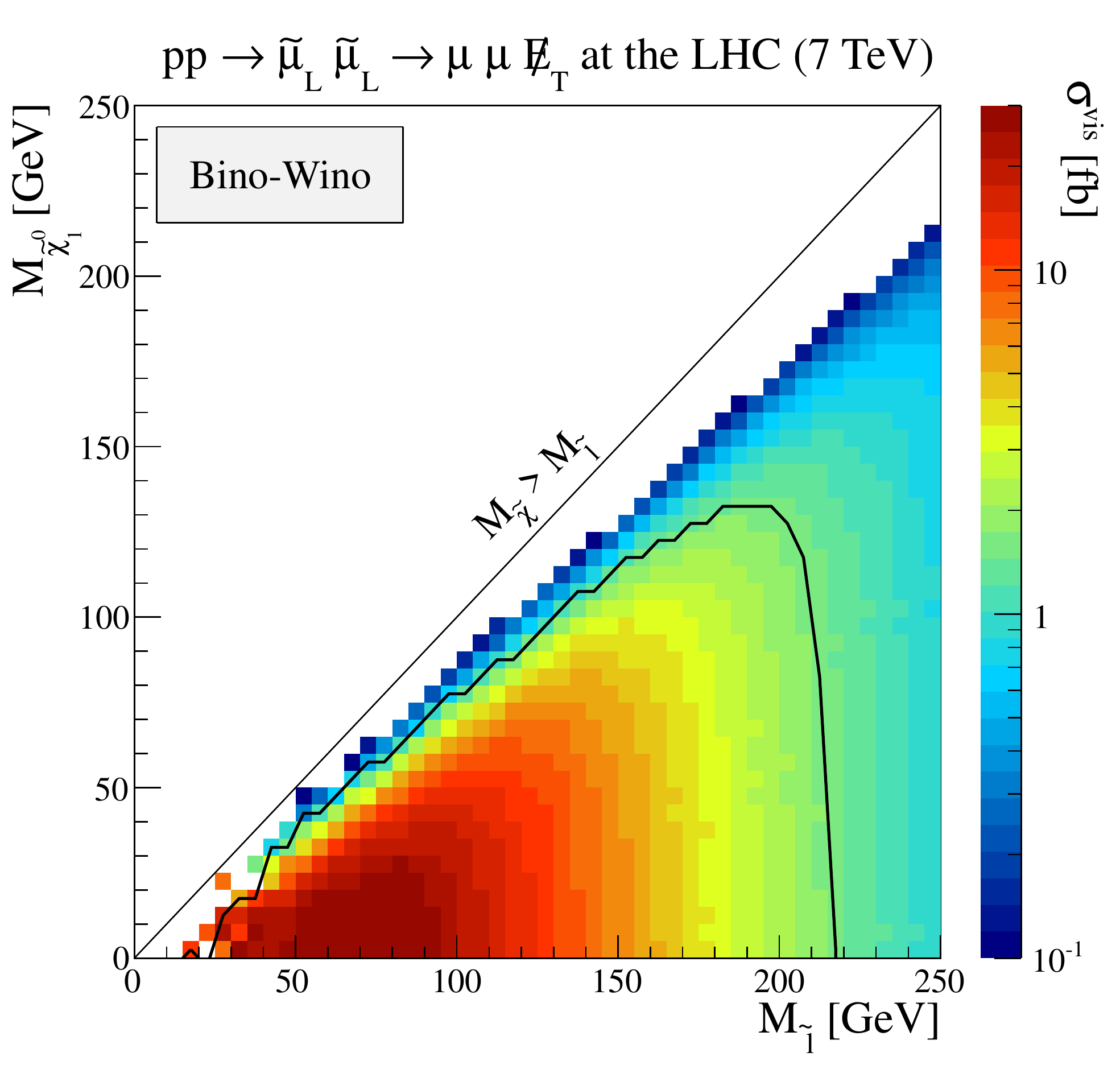}
  \includegraphics[width=\columnwidth]{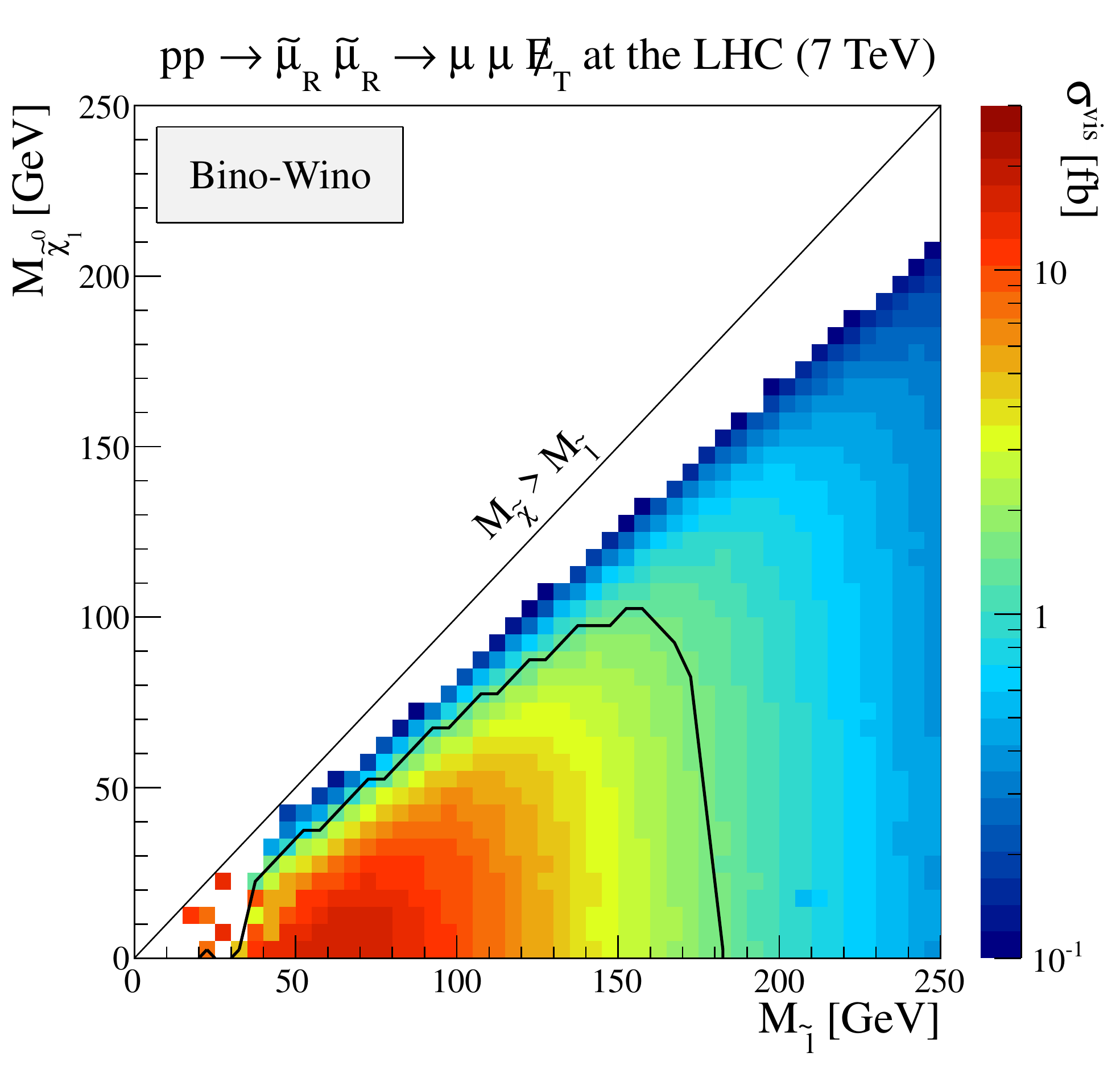}
\caption{\label{fig:5}95\% confidence exclusion limit for left-handed
(top) and right-handed (bottom) smuon pair production,
given in the $(M_{\tilde{l}},M_{\tilde\chi^0_1})$ mass plane of a simplified
model for mixed bino-wino neutralino nature.
We present the visible cross
section after applying the ATLAS selection strategy.
The limits are extracted for 4.7~fb$^{-1}$ of LHC collisions at a centre-of-mass
energy of 7~TeV.}
\end{figure}

The situation is quite different for the left-/right-handed nature of the
decaying slepton, as one can see in Fig.\ \ref{fig:5}. Here we assume a mixed
%
%%%%%%%%%%%%%% Begin Figure 5 %%%%%%%%%%%%%%%%%%%%%%%%%%%%%%%%%%%%%%%%%%
%%%%%%%%%%%%%% End of Figure 5 %%%%%%%%%%%%%%%%%%%%%%%%%%%%%%%%%%%%%%%%%
%
bino-wino nature for the lightest neutralino and study the production of
left- (top) and right-handed (bottom) smuons. The exclusion curves are in
this case quite different, reflecting the fact that right-handed (s)leptons
have weaker couplings and smaller cross sections.

\section{Conclusion}
\label{}

The most precise electroweak SUSY particle production cross sections at NLO+NLL
are by now routinely taken into account by the LHC experiments ATLAS and CMS
for gaugino/higgsino and slepton searches, in particular when deriving exclusion
limits. The corresponding computer code RESUMMINO has been made public and is
available for use in experimental analyses and theoretical studies, not only
for SUSY particles \cite{Fuks:2013vua}, but also for additional neutral gauge
bosons \cite{Fuks:2007gk}.



%% References

\nocite{*}
\bibliographystyle{elsarticle-num}
\bibliography{martin}

\begin{thebibliography}{00}



%\cite{Beenakker:1996ch}
\bibitem{Beenakker:1996ch}
  W.~Beenakker, R.~Hopker, M.~Spira and P.~M.~Zerwas,
  %``Squark and gluino production at hadron colliders,''
  Nucl.\ Phys.\ B {\bf 492} (1997) 51
  [hep-ph/9610490].
  %%CITATION = HEP-PH/9610490;%%
  %683 citations counted in INSPIRE as of 22 Jul 2014

%\cite{Beenakker:1997ut}
\bibitem{Beenakker:1997ut}
  W.~Beenakker, M.~Kramer, T.~Plehn, M.~Spira and P.~M.~Zerwas,
  %``Stop production at hadron colliders,''
  Nucl.\ Phys.\ B {\bf 515} (1998) 3
  [hep-ph/9710451].
  %%CITATION = HEP-PH/9710451;%%
  %350 citations counted in INSPIRE as of 22 Jul 2014

%\cite{Berger:1998kh}
\bibitem{Berger:1998kh}
  E.~L.~Berger, M.~Klasen and T.~M.~P.~Tait,
  %``Scale dependence of squark and gluino production cross-sections,''
  Phys.\ Rev.\ D {\bf 59} (1999) 074024
  [hep-ph/9807230].
  %%CITATION = HEP-PH/9807230;%%
  %19 citations counted in INSPIRE as of 17 Jul 2014

%\cite{Beenakker:1999xh}
\bibitem{Beenakker:1999xh}
  W.~Beenakker, M.~Klasen, M.~Kr\"amer, T.~Plehn, M.~Spira and P.~M.~Zerwas,
  %``The Production of charginos / neutralinos and sleptons at hadron colliders,''
  Phys.\ Rev.\ Lett.\  {\bf 83} (1999) 3780
   [Erratum-ibid.\  {\bf 100} (2008) 029901]
  [hep-ph/9906298].
  %%CITATION = HEP-PH/9906298;%%
  %307 citations counted in INSPIRE as of 17 Jul 2014

%\cite{Berger:1999mc}
\bibitem{Berger:1999mc}
  E.~L.~Berger, M.~Klasen and T.~M.~P.~Tait,
  %``Associated production of gauginos and gluinos at hadron colliders in next-to-leading order SUSY QCD,''
  Phys.\ Lett.\ B {\bf 459} (1999) 165
  [hep-ph/9902350].
  %%CITATION = HEP-PH/9902350;%%
  %25 citations counted in INSPIRE as of 17 Jul 2014

%\cite{Berger:2000iu}
\bibitem{Berger:2000iu}
  E.~L.~Berger, M.~Klasen and T.~M.~P.~Tait,
  %``Next-to-leading order SUSY QCD predictions for associated production of gauginos and gluinos,''
  Phys.\ Rev.\ D {\bf 62} (2000) 095014
  [hep-ph/0005196]
  %%CITATION = HEP-PH/0005196;%%
  %33 citations counted in INSPIRE as of 17 Jul 2014
and
%\cite{Berger:2002vd}
%\bibitem{Berger:2002vd}
%  E.~L.~Berger, M.~Klasen and T.~M.~P.~Tait,
  %``Erratum: Next-to-leading order supersymmetric QCD predictions for associated production of gauginos and gluinos (Phys.Rev.D62:095014,2000),''
  Phys.\ Rev.\ D {\bf 67} (2003) 099901
  [hep-ph/0212306].
  %%CITATION = HEP-PH/0212306;%%
  %12 citations counted in INSPIRE as of 17 Jul 2014

%\cite{Spira:2002rd}
\bibitem{Spira:2002rd}
  M.~Spira,
  %``Higgs and SUSY particle production at hadron colliders,''
  hep-ph/0211145.
  %%CITATION = HEP-PH/0211145;%%
  %55 citations counted in INSPIRE as of 17 Jul 2014

%\cite{Kramer:2012bx}
\bibitem{Kramer:2012bx}
  M.~Kramer, A.~Kulesza, R.~van der Leeuw, M.~Mangano, S.~Padhi, T.~Plehn and X.~Portell,
  %``Supersymmetry production cross sections in $pp$ collisions at $\sqrt{s}=7$ TeV,''
  arXiv:1206.2892 [hep-ph]
  %%CITATION = ARXIV:1206.2892;%%
  %195 citations counted in INSPIRE as of 22 Jul 2014
and references therein.

%\cite{Debove:2009ia}
\bibitem{Debove:2009ia}
  J.~Debove, B.~Fuks and M.~Klasen,
  %``Transverse-momentum resummation for gaugino-pair production at hadron colliders,''
  Phys.\ Lett.\ B {\bf 688} (2010) 208
  [arXiv:0907.1105 [hep-ph]].
  %%CITATION = ARXIV:0907.1105;%%
  %17 citations counted in INSPIRE as of 17 Jul 2014

%\cite{Debove:2010kf}
\bibitem{Debove:2010kf}
  J.~Debove, B.~Fuks and M.~Klasen,
  %``Threshold resummation for gaugino pair production at hadron colliders,''
  Nucl.\ Phys.\ B {\bf 842} (2011) 51
  [arXiv:1005.2909 [hep-ph]].
  %%CITATION = ARXIV:1005.2909;%%
  %12 citations counted in INSPIRE as of 17 Jul 2014

%\cite{Debove:2011xj}
\bibitem{Debove:2011xj}
  J.~Debove, B.~Fuks and M.~Klasen,
  %``Joint Resummation for Gaugino Pair Production at Hadron Colliders,''
  Nucl.\ Phys.\ B {\bf 849} (2011) 64
  [arXiv:1102.4422 [hep-ph]].
  %%CITATION = ARXIV:1102.4422;%%
  %13 citations counted in INSPIRE as of 17 Jul 2014

%\cite{Fuks:2012qx}
\bibitem{Fuks:2012qx}
  B.~Fuks, M.~Klasen, D.~R.~Lamprea and M.~Rothering,
  %``Gaugino production in proton-proton collisions at a center-of-mass energy of 8 TeV,''
  JHEP {\bf 1210} (2012) 081
  [arXiv:1207.2159 [hep-ph]].
  %%CITATION = ARXIV:1207.2159;%%
  %16 citations counted in INSPIRE as of 17 Jul 2014

%\cite{Bozzi:2006fw}
\bibitem{Bozzi:2006fw}
  G.~Bozzi, B.~Fuks and M.~Klasen,
  %``Transverse-momentum resummation for slepton-pair production at the CERN LHC,''
  Phys.\ Rev.\ D {\bf 74} (2006) 015001
  [hep-ph/0603074].
  %%CITATION = HEP-PH/0603074;%%
  %35 citations counted in INSPIRE as of 17 Jul 2014

%\cite{Bozzi:2007qr}
\bibitem{Bozzi:2007qr}
  G.~Bozzi, B.~Fuks and M.~Klasen,
  %``Threshold Resummation for Slepton-Pair Production at Hadron Colliders,''
  Nucl.\ Phys.\ B {\bf 777} (2007) 157
  [hep-ph/0701202].
  %%CITATION = HEP-PH/0701202;%%
  %31 citations counted in INSPIRE as of 17 Jul 2014

%\cite{Bozzi:2007tea}
\bibitem{Bozzi:2007tea}
  G.~Bozzi, B.~Fuks and M.~Klasen,
  %``Joint resummation for slepton pair production at hadron colliders,''
  Nucl.\ Phys.\ B {\bf 794} (2008) 46
  [arXiv:0709.3057 [hep-ph]].
  %%CITATION = ARXIV:0709.3057;%%
  %31 citations counted in INSPIRE as of 17 Jul 2014

%\cite{Fuks:2013lya}
\bibitem{Fuks:2013lya}
  B.~Fuks, M.~Klasen, D.~R.~Lamprea and M.~Rothering,
  %``Revisiting slepton pair production at the Large Hadron Collider,''
  JHEP {\bf 1401} (2014) 168
  [arXiv:1310.2621, arXiv:1310.2621 [hep-ph]].
  %%CITATION = ARXIV:1310.2621,;%%
  %9 citations counted in INSPIRE as of 17 Jul 2014

%\cite{Kramer:1996iq}
\bibitem{Kramer:1996iq}
  M.~Kramer, E.~Laenen and M.~Spira,
  %``Soft gluon radiation in Higgs boson production at the LHC,''
  Nucl.\ Phys.\ B {\bf 511} (1998) 523
  [hep-ph/9611272].
  %%CITATION = HEP-PH/9611272;%%
  %251 citations counted in INSPIRE as of 29 Jul 2014

%\cite{Alwall:2011uj}
\bibitem{Alwall:2011uj}
  J.~Alwall, M.~Herquet, F.~Maltoni, O.~Mattelaer and T.~Stelzer,
  %``MadGraph 5 : Going Beyond,''
  JHEP {\bf 1106} (2011) 128
  [arXiv:1106.0522 [hep-ph]].
  %%CITATION = ARXIV:1106.0522;%%
  %1254 citations counted in INSPIRE as of 28 Jul 2014

%\cite{Sjostrand:2006za}
\bibitem{Sjostrand:2006za}
  T.~Sjostrand, S.~Mrenna and P.~Z.~Skands,
  %``PYTHIA 6.4 Physics and Manual,''
  JHEP {\bf 0605} (2006) 026
  [hep-ph/0603175].
  %%CITATION = HEP-PH/0603175;%%
  %5355 citations counted in INSPIRE as of 28 Jul 2014

%\cite{Mangano:2002ea}
\bibitem{Mangano:2002ea}
  M.~L.~Mangano, M.~Moretti, F.~Piccinini, R.~Pittau and A.~D.~Polosa,
  %``ALPGEN, a generator for hard multiparton processes in hadronic collisions,''
  JHEP {\bf 0307} (2003) 001
  [hep-ph/0206293].
  %%CITATION = HEP-PH/0206293;%%
  %2218 citations counted in INSPIRE as of 29 Jul 2014

%\cite{Conte:2012fm}
\bibitem{Conte:2012fm}
  E.~Conte, B.~Fuks and G.~Serret,
  %``MadAnalysis 5, A User-Friendly Framework for Collider Phenomenology,''
  Comput.\ Phys.\ Commun.\  {\bf 184} (2013) 222
  [arXiv:1206.1599 [hep-ph]].
  %%CITATION = ARXIV:1206.1599;%%
  %39 citations counted in INSPIRE as of 29 Jul 2014

%\cite{Bozzi:2004qq}
\bibitem{Bozzi:2004qq}
  G.~Bozzi, B.~Fuks and M.~Klasen,
  %``Slepton production in polarized hadron collisions,''
  Phys.\ Lett.\ B {\bf 609} (2005) 339
  [hep-ph/0411318].
  %%CITATION = HEP-PH/0411318;%%
  %25 citations counted in INSPIRE as of 17 Jul 2014

%\cite{Debove:2008nr}
\bibitem{Debove:2008nr}
  J.~Debove, B.~Fuks and M.~Klasen,
  %``Model-independent analysis of gaugino-pair production in polarized and unpolarized hadron collisions,''
  Phys.\ Rev.\ D {\bf 78} (2008) 074020
  [arXiv:0804.0423 [hep-ph]].
  %%CITATION = ARXIV:0804.0423;%%
  %16 citations counted in INSPIRE as of 17 Jul 2014

%\cite{Aad:2012pxa}
\bibitem{Aad:2012pxa}
  G.~Aad {\it et al.}  [ATLAS Collaboration],
  %``Search for direct slepton and gaugino production in final states with two leptons and missing transverse momentum with the ATLAS detector in $pp$ collisions at $\sqrt{s}=7$ TeV,''
  Phys.\ Lett.\ B {\bf 718} (2013) 879
  [arXiv:1208.2884 [hep-ex]].
  %%CITATION = ARXIV:1208.2884;%%
  %62 citations counted in INSPIRE as of 28 Jul 2014

%\cite{Fuks:2013vua}
\bibitem{Fuks:2013vua}
  B.~Fuks, M.~Klasen, D.~R.~Lamprea and M.~Rothering,
  %``Precision predictions for electroweak superpartner production at hadron colliders with Resummino,''
  Eur.\ Phys.\ J.\ C {\bf 73} (2013) 2480
  [arXiv:1304.0790 [hep-ph]].
  %%CITATION = ARXIV:1304.0790;%%
  %13 citations counted in INSPIRE as of 17 Jul 2014

%\cite{Fuks:2007gk}
\bibitem{Fuks:2007gk}
  B.~Fuks, M.~Klasen, F.~Ledroit, Q.~Li and J.~Morel,
  %``Precision predictions for $Z^\prime$ - production at the CERN LHC: QCD matrix elements, parton showers, and joint resummation,''
  Nucl.\ Phys.\ B {\bf 797} (2008) 322
  [arXiv:0711.0749 [hep-ph]].
  %%CITATION = ARXIV:0711.0749;%%
  %38 citations counted in INSPIRE as of 17 Jul 2014

\end{thebibliography}



\end{document}